\documentclass{iaus}
\usepackage{natbib}
\usepackage{graphicx}
\newcommand{\aj}{\textit{AJ}} 				%%%%%             Astronomical Journal
\newcommand{\apj}{\textit{ApJ}} 			%%%%%             Astrophysical Journal
\newcommand{\apjl}{\textit{ApJ}} 			%%%%%             Astrophysical Journal
\newcommand{\apjs}{\textit{ApJS}} 			%%%%%             Astrophysical Journal Supplement Series
\newcommand{\apss}{\textit{Ap\&SS}} 		%%%%%             Astrophysics and Space Science
\newcommand{\aap}{\textit{A\&A}} 			%%%%%             Astronomy & Astrophysics
\newcommand{\mnras}{\textit{MNRAS}} 		%%%%%             Monthly Notices of the RAS
\newcommand{\qjras}{\textit{QJRAS}} 		%%%%%             Quarterly Journal of the RAS
\title[HB Stars and The Age Debate] %% give here short title %%
{The Ages of Stars: The Horizontal Branch}

\author[M. Catelan]   %% give here short author list %%
{M. Catelan$^1$
  \thanks{John Simon Guggenheim Memorial Foundation Fellow.}
  \thanks{On sabbatical leave at Catholic University of America, 
Dept. of Physics, Washington, DC.
}
}

\affiliation{$^1$Pontificia Universidad Cat\'olica de Chile,
Departamento de Astronom\'{i}a y Astrof\'{i}sica, \\ Av. Vicu\~na 
Mackenna 4860, 782-0436 Macul, Santiago, Chile \\email: {\tt mcatelan@astro.puc.cl}}

\pubyear{2008}
\volume{258}  %% insert here IAU Symposium No.
\pagerange{100--109}
\date{\today}
\jname{The Ages of Stars}
\editors{D. Soderblom}
\begin{document}

\maketitle

\begin{abstract}

Horizontal branch (HB) stars play a particularly important role in the ``age debate,'' since they are at the very center of the long-standing ``second parameter'' problem. In this review, I discuss some recent progress in our understanding of the nature and origin of HB stars.

\end{abstract}

\firstsection % if your document starts with a section,

\section{Introduction}
Horizontal branch (HB) stars have long played a central role in the 
age debate. These low-mass stars, which burn helium in the core and 
hydrogen in a shell, are the immediate progeny of the luminous, 
vigorously mass-losing red giant branch (RGB) stars. Most importantly 
in the present context, their temperatures depend strongly on their 
total masses. More specifically, the 
lower the mass of an HB star, the bluer it becomes, by the time it 
reaches the zero-age HB (ZAHB). Therefore, the HB morphology in 
globular clusters (GC's) is naturally expected to become bluer with age. 

It has long been known that the {\em first parameter} 
controlling HB morphology is actually metallicity, with 
more metal-rich GC's presenting redder HB's than 
their more metal-poor counterparts. Still, \citet{sw60} first 
realized, based mainly on the early 
observations of the GC's M3 (NGC~5272), M13 (NGC~6205),
and M22 (NGC~6656) by \citet{as53}, \citet{aj55}, and \citet{am59}, 
that GC's with a {\em given} 
metallicity might also present widely different HB types, due to the 
action of an unknown ``second parameter.'' We quote
from their study: 

\begin{quotation}
``... the character of the horizontal branch is spoiled by the two clusters M13 and M22. (...) M13 appear[s] to be metal-rich, whereas the character of the horizontal branch simulates that of the very weak-lined group (M15, M92, NGC 5897). (...) M13 is younger than M2 or M5 (...) Consequently, in addition to chemical composition, the second parameter of age may be affecting the correlations.''
\end{quotation}

\noindent (Note that the sense of the correlation between age and HB 
morphology suggested by \citeauthor{sw60} is the {\em opposite} of 
what modern studies indicate to be necessary to account for the 
second-parameter phenomenon.) 
 
It soon became clear that age was not the only possible 
second parameter candidate. By the early 1970's, the list of 
candidates had increased sharply, and already included, in addition to 
age, the helium abundance and the abundances of the CNO elements
\citep{rtr73}. While the ``age as the second parameter'' 
scenario was to gain an important boost with the work by \citet{sz78}, 
who noted that HB morphology tends to become redder with 
increasing Galactocentric distance~-- which was interpreted as an age
effect, with more distant clusters being younger on average, and having
possibly been accreted from ``protogalactic fragments'' of external origin
over the Galaxy's lifetime~-- many other second parameter candidates 
have also surfaced over the years. This includes, among others, 
cluster concentration, total mass, and ellipticity; stellar rotation; 
magnetic fields; planetary systems; and mass loss on the RGB 
\citep[see][for extensive references]{mc08}.   

\begin{figure}[!tp]
  \centering
  {\includegraphics[width=5.275in]{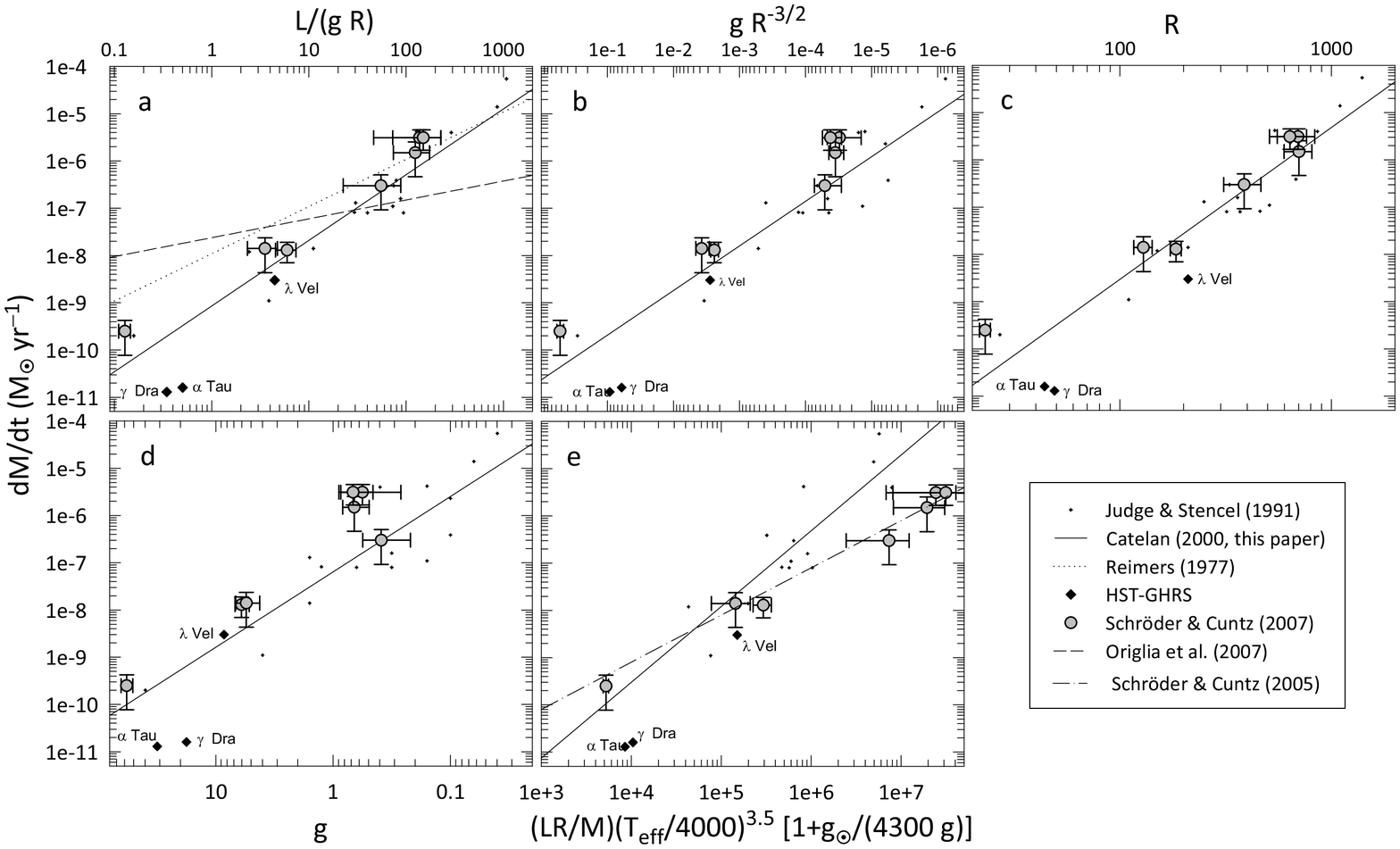}} 
  \caption{Comparison between different mass loss recipes and the empirical data. The 
  small crosses represent data from \cite{js91}, selected according to \citet{mc00}. 
  Filled diamonds correspond to the HST-GHRS results by \citet{rrea98} and 
  \citet{dmea98}. Gray symbols with error bars represent data from the recent 
  compilation by \citet{sc07}. In all panels, the {\em solid lines} show the fits
  derived by \citet{mc00} from the \citeauthor{js91} data (using the different 
  combinations of physical parameters indicated in the $x$-axis of each plot 
  as the independent variable), except for panel {\em e}, 
  where the fit is presented here for the first time. 
  In panel {\em a}, the {\em dotted line} represents the predicted mass loss rates 
  according to the \citet{dr75} formula, whereas the {\em dashed line} indicates the 
  predicted mass loss rates according to \citet{loea07}. In panel {\em e}, the 
  {\em dash-dotted line} indicates the mass loss rates predicted by the \citet{sc05}
  formula.}    
  \label{fig:SC-DM}
\end{figure}

\begin{figure}[!tp]
  \centering
%  \scalebox{0.3}
  \includegraphics[width=5.325in]{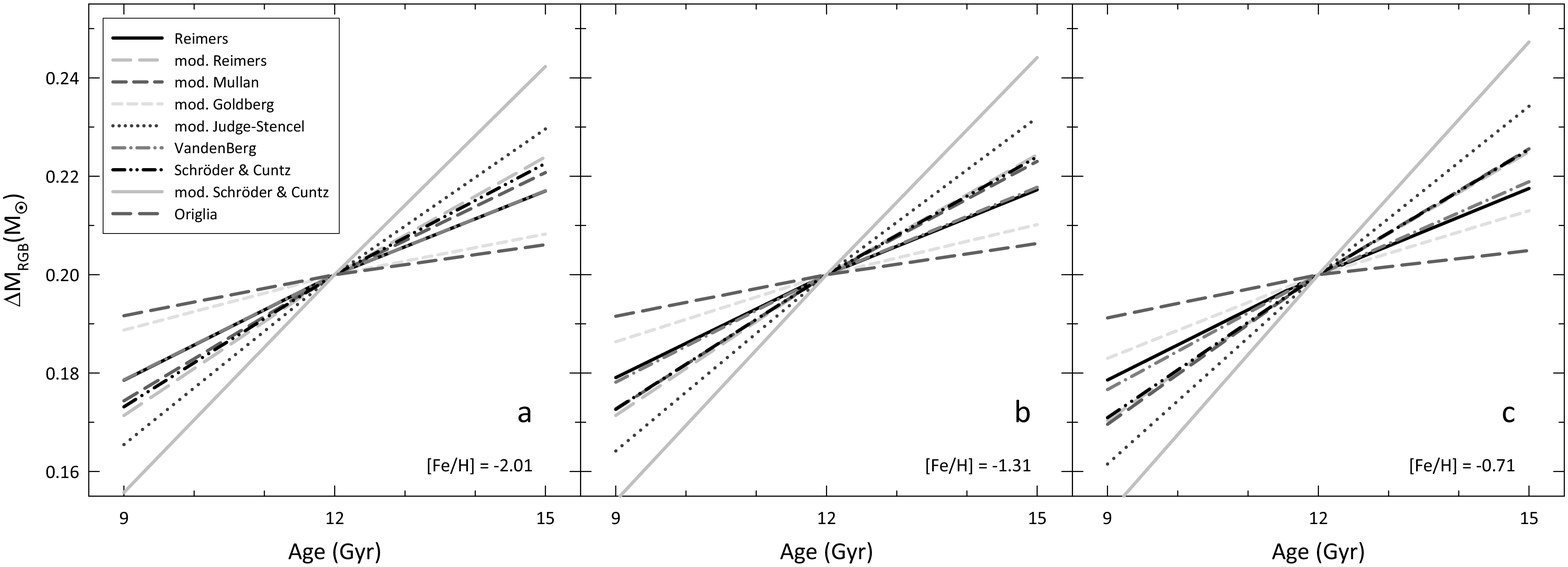}
  \vskip 0.5cm
%  \scalebox{0.3}
  \includegraphics[width=5.325in]{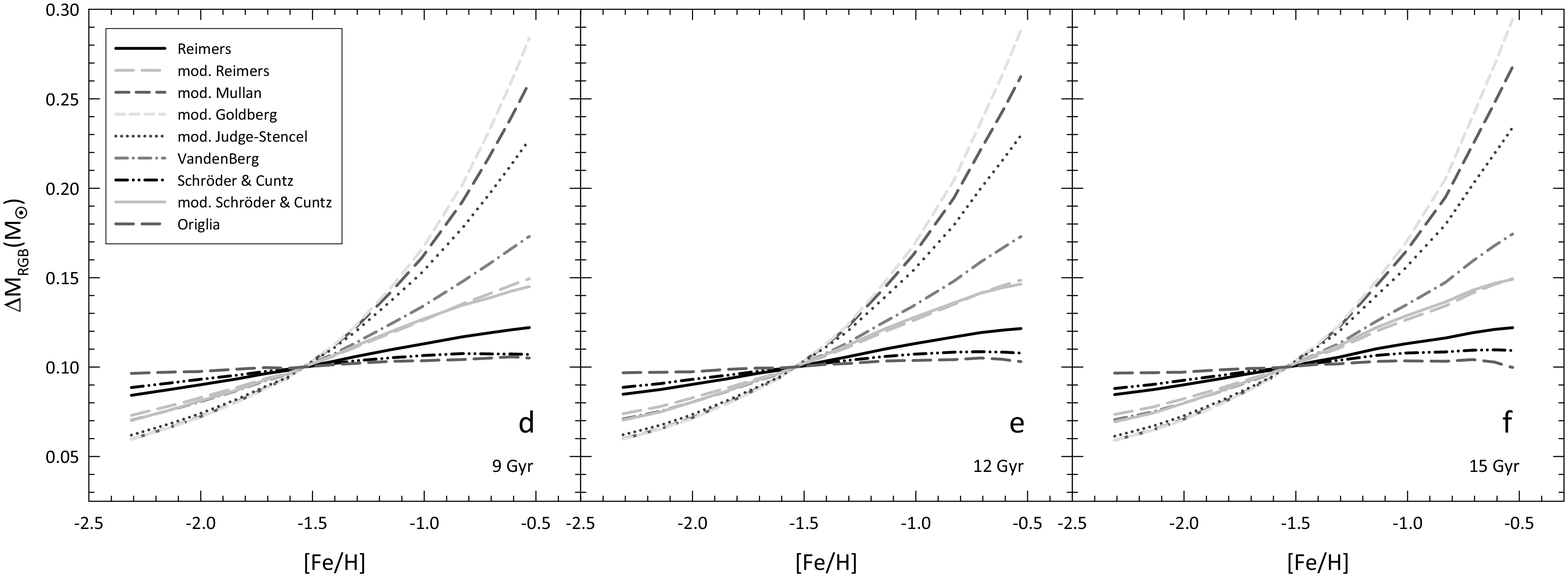}
  \caption{Dependence of the integrated RGB mass loss upon age for fixed metallicity
    (panel {\em a}: ${\rm [Fe/H]} = -2.01$; panel {\em b}: ${\rm [Fe/H]} = -1.31$;    
	panel {\em c}: ${\rm [Fe/H]} = -0.71$) and upon metallicity for a fixed age
    (panel {\em d}: $9$~Gyr;  panel {\em e}: $12$~Gyr;  panel {\em f}: 
	$15$~Gyr), for the different mass loss recipes indicated. In panels 
	{\em a} through {\em c}, the total mass loss has been normalized to a value
	of $0.20\, M_{\odot}$ at 12~Gyr; in panels {\em d} through {\em f}, in turn, 
	the integrated mass loss has been normalized to a value of $0.10\, M_{\odot}$
	at ${\rm [Fe/H]} = -1.54$.} 
  \label{fig:DM-Age-FeH}
\end{figure}

\begin{figure}[ht]
\begin{center}
  \centerline{
 \includegraphics[width=4.575in]{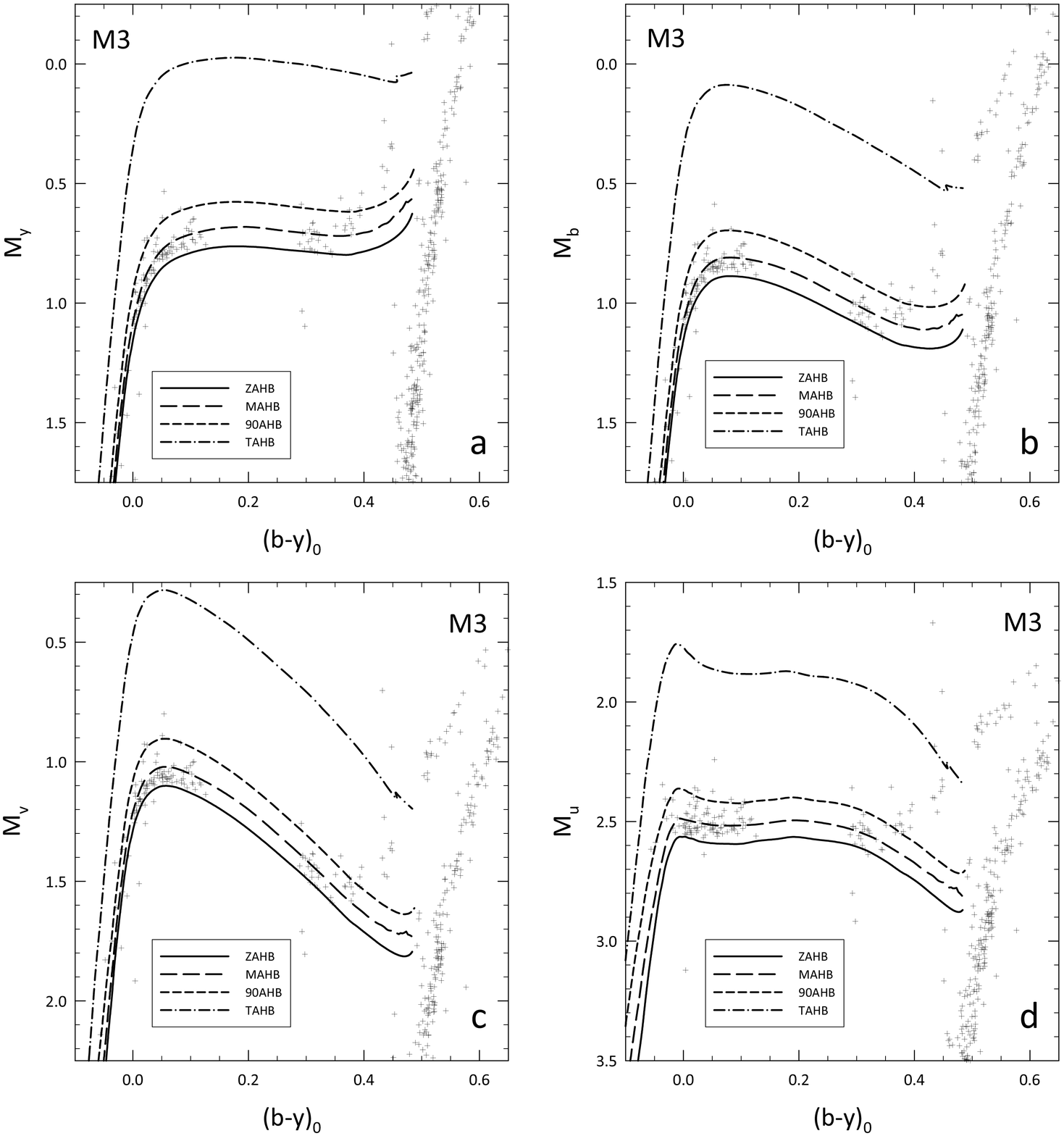} 
 }
 \caption{Comparison between \citet{bs63} photometry for M3, as derived by 
   \citet{fgea98,fgea99}, with the predictions of theoretical models for  
   $Y_{\rm MS} = 0.23$, $Z = 0.002$. The observed data were shifted
   vertically so as to lead to a good match with the theoretical ZAHB
   at the red HB.
   %The reference loci have the following
   %meanings: ZAHB~= zero-age HB; MAHB~= middle-age HB; 90AHB~= 90\%-age HB; 
   %TAHB~= terminal-age HB (see \S\ref{sec:M3} for further details).
   }
   \label{fig:M3}
\end{center}
\end{figure}

\begin{figure}[t]
\begin{center}
  \centerline{
 \includegraphics[width=5.325in]{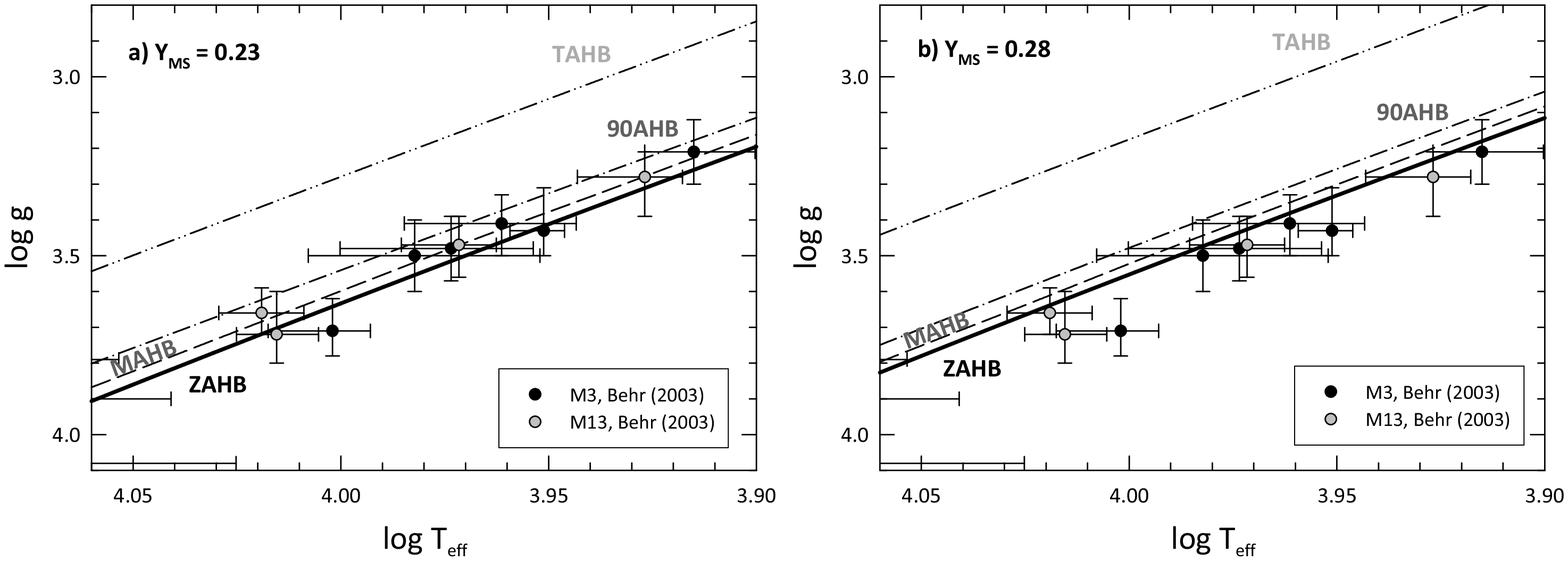} 
 }
 \caption{Comparison between spectroscopically derived gravities for blue 
   HB stars in M3 and M13 \citep[from][]{bb03} with the same theoretical 
   models as before, but for two different $Y_{\rm MS}$ values: 23\% 
   (panel {\em a}) and 28\% (panel {\em b}).}
   \label{fig:M3-logg}
\end{center}
\end{figure}

While it seems clear now that age does play an important role, it has 
also become evident that it is 
not the only parameter involved. Indeed, the 
presence of bimodal HB's in such GC's as NGC~2808 have long 
pointed to the need for other second parameters in addition to age  
\citep[e.g.,][]{rtrea93}. Recent, deep CMD studies 
have revealed that some of the most massive globulars, 
NGC~2808 included, present a surprisingly complex history of star 
formation, with the presence of extreme levels of helium enhancement 
among at least some of their stars 
\citep[e.g.,][]{jn04,fdea05,gpea05,gpea07}. As noted by these authors, 
such He enhancement would provide a natural explanation for the 
presence of hot HB stars in these clusters. High helium abundances 
also appear to provide a natural explanation for some of the  
peculiarities observed in the CMD's and RR Lyrae properties
in the GC's NGC~6388 and NGC~6441 
\citep[e.g.,][and references therein]{mcea06,cd07}.

In the next few sections, we will address some empirical constraints 
that may be posed on some of these second parameter candidates.

\section{Mass Loss in Red Giants}\label{sec:DM} 
In order to reliably predict the temperature of an HB star 
of a given composition and age, we must know how much mass it 
loses on the RGB. 
Unfortunately, our knowledge of RGB mass loss remains rather limited. 
Most studies adopt the \citet{dr75} mass loss formula to predict 
the integrated mass loss along the RGB. However, 
recent evidence indicates that the \citeauthor{dr75} formula is
not a reliable description of RGB mass loss  
\citep[e.g.,][]{mc00,mc08,sc05,sc07}. In addition, there are several 
alternative mass loss formulations which may better 
describe the available data. We illustrate this point by 
comparing, in Figure~\ref{fig:SC-DM}, some empirical mass loss rates 
with the predicted rates from several of these alternative 
mass loss formulae \citep[see][for extensive references]{mc00}. While
the \citeauthor{dr75} formula is clearly inconsistent with the data, 
the empirical data cannot conclusively distinguish among these
alternative formulations.\footnote{Note that, while the \citet{loea07} 
mass loss formula is inconsistent with the plotted data, this should 
not be taken as evidence against its validity, since this formula 
has been suggested to apply exclusively to low-metallicity stars. 
Still, some caveats regarding the 
\citeauthor{loea07} study have recently been raised 
\citep[see][]{mbea08}.}  

The serious problem which this uncertainty in the RGB mass loss poses 
for our understanding of HB 
morphology and its dependence on age and metallicity 
is apparent from Figure~\ref{fig:DM-Age-FeH}, where the 
integrated RGB mass loss is plotted as a function of the 
age for fixed metallicity (panels {\em a} through {\em c}) and as a 
function of metallicity for fixed age (panels {\em d} through {\em f}). 
More specifically, we know that only a very mild 
$\Delta M - {\rm [Fe/H]}$ dependence can account for
the observed relation between HB type and [Fe/H] without 
resorting to a significant age-metallicity relation 
\citep[see, e.g., Fig.~1a in][]{ldz94}. In this sense, we find that 
the \citet{loea07} and \citet{sc05} mass loss formulae lead to the 
weakest $\Delta M - {\rm [Fe/H]}$ dependence. 
All other formulae that we have tested lead 
to steeper dependencies between $\Delta M$ and [Fe/H] than the 
\citet{dr75} relation, thus implying steeper dependencies between 
age and metallicity as well. The precise dependence 
between
$\Delta M$ and metallicity is also important in terms of explaining
the ultraviolet upturn phenomenon of elliptical galaxies and spiral
bulges \citep[see][for a recent review]{mc07}.

Similarly, a stronger dependence between $\Delta M$ and age at fixed
[Fe/H] makes it easier to account for a given 
second-parameter pair in terms solely of an age difference. 
According to Figure~\ref{fig:DM-Age-FeH}, the equation that is most
successful in this regard is a modified version of the \citet{sc05}
formula, in which the adopted power law exponents are obtained 
by a least-squares fit to the \citet{js91} data, selected as in 
\citet{mc00} (see Fig.~\ref{fig:SC-DM}{\em e}). While not the steepest, 
the original \citeauthor{sc05} formula provides a stronger dependence
between $\Delta M$ and age than does the \citet{dr75} formula, which 
should reduce the required age difference between second parameter
pairs. By contrast, 
the \citet{loea07} equation shows the weakest dependence, 
with a remarkably constant integrated $\Delta M$ value over a wide 
range in ages.

\section{Helium Enrichment in Globular Clusters}
As previously noted (\S1), high levels of helium enrichment have 
been detected among some of the most massive Galactic GC's. 
Very recently, it has been suggested that such helium enhancements 
are in fact not the exception, but indeed the rule, among Galactic 
GC's \citep{dac08}. Here we provide a first test of 
this scenario, in the case of the GC's M3 and NGC~1851. 

\subsection{The Case of M3}\label{sec:M3}
\citet{dac08} and \citet{cd08} claim that the blue HB component 
in M3 owes its origin to a moderate level of He enhancement in the 
cluster, between 2\% and 6\%. Is this supported by the available data? 

To answer this question, we compare, in Figure~\ref{fig:M3}, canonical 
theoretical predictions from \citet{mcea98} and \citet{sc98} for a 
helium abundance of $Y_{\rm MS} = 0.23$ and a metallicity 
$Z = 0.002$ with high-precision photometry, in the \citet{bs63} system, 
from \citet{fgea98,fgea99}. The empirical data were corrected for 
reddening following \citet{wh96}. 
In these plots, the drawn lines represent
different fiducial loci, as follows: ZAHB, middle-age HB (MAHB, or 
average locus occupied by the HB stars), 90\%-age HB (90AHB, or locus
below which one should expect to find about 90\% of all HB stars), 
and terminal-age HB (TAHB, or He exhaustion locus). Except for a 
discrepancy between the predicted and observed numbers of highly 
evolved stars (both on the blue {\em and} red HB sides of the 
distribution) that was previously noted by \citet{vc08}, one finds 
remarkable agreement between the model predictions for a constant $Y$ 
and the observations. Such an agreement is also confirmed by the 
spectroscopic data from \citet{bb03}, as can be seen from 
Figure~\ref{fig:M3-logg}, where we limit the comparison to temperatures 
lower than 11,500~K due to the well-known complications brought about 
by the ``Grundahl jump'' phenomenon \citep{fgea99}. Interestingly, this
plot also appears to support a similar helium abundance between  
M3's blue HB stars and the redder blue HB stars in M13.

\begin{figure}[t]
\begin{center}
  \centerline{
 \includegraphics[width=5.325in]{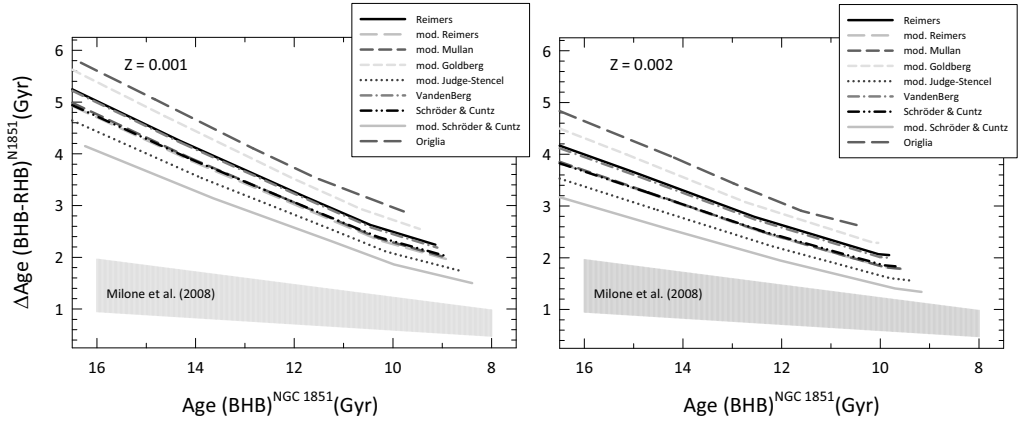} 
 }
 \caption{The age difference that is required to account for the difference 
   in HB morphology between the blue and red components of NGC~1851's HB
   ({\em lines}, based on the different mass loss formulae discussed in 
   \S\ref{sec:DM}) is compared with the age difference that is estimated 
   from the observed split on the SGB ({\em gray band}).   }
   \label{fig:N1851-ages}
\end{center}
\end{figure}

\begin{figure}[t]
\begin{center}
  \centerline{
 \includegraphics[width=4.575in]{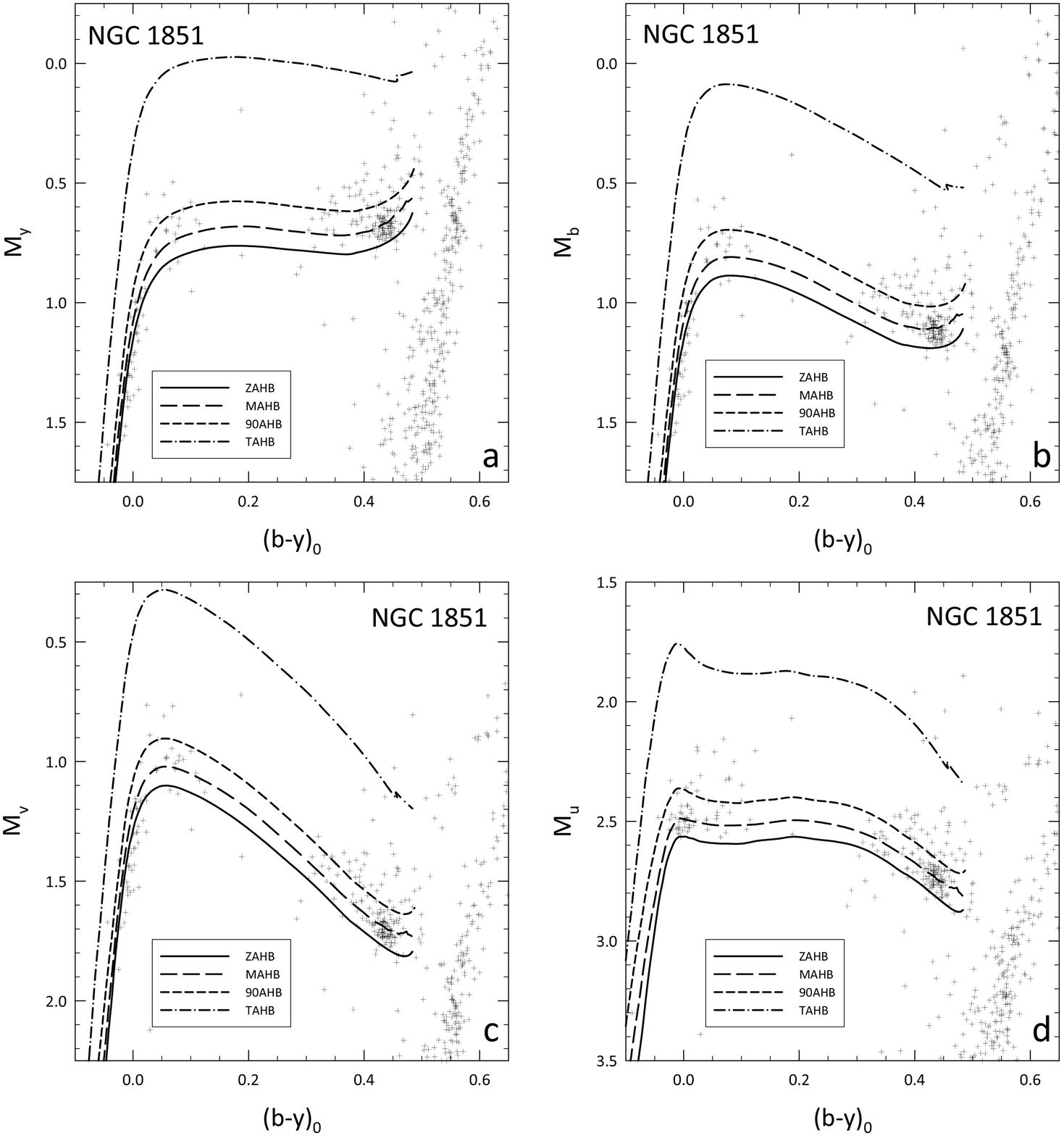} 
 }
 \caption{As in Figure~\ref{fig:M3}, but for NGC~1851.}
   \label{fig:N1851}
\end{center}
\end{figure}

\begin{figure}[t]
\begin{center}
  \centerline{
 \includegraphics[width=4.575in]{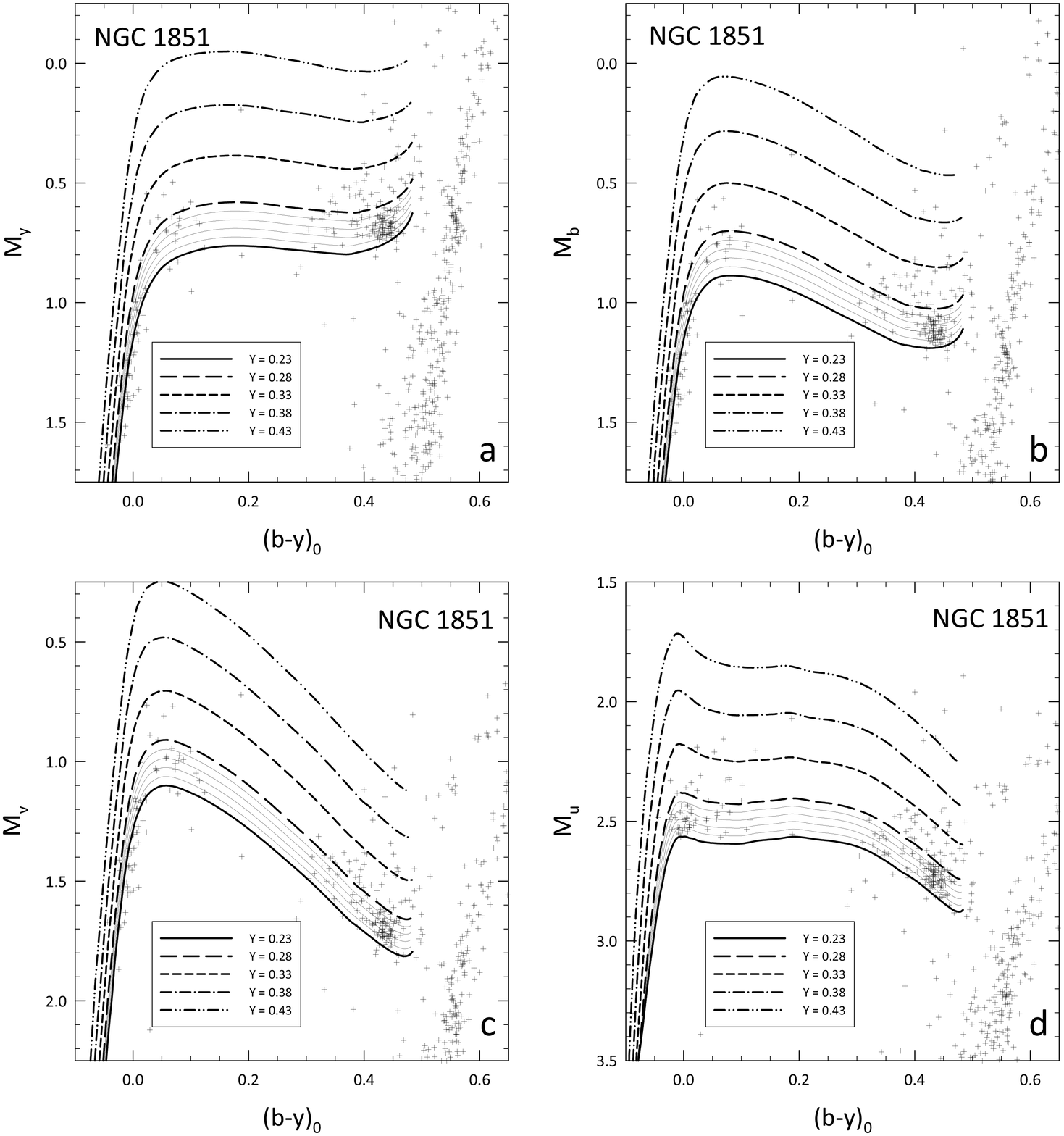} 
 }
 \caption{As in Figure~\ref{fig:N1851}, but showing ZAHB's for the several 
   different indicated $Y_{\rm MS}$ values. Interpolated ZAHB's are also 
   plotted between the 23\% and 28\% loci, at intervals of 1\%.}
   \label{fig:N1851-Y}
\end{center}
\end{figure}

\subsection{The Case of NGC~1851}
\citet{apmea08} have recently discovered that the subgiant branch (SGB) 
of NGC~1851 is actually split into two separate components, which may 
be linked to the cluster's well-known bimodal HB morphology. The 
most straightforward explanation for this split would be a difference
in age by $1.0 \pm 0.4$~Gyr. However, as shown in Figure~\ref{fig:N1851-ages}, 
this is inconsistent with the age 
difference that would be required to fully account for the separation 
between the blue and red HB components of the cluster, irrespective 
of the mass loss formula (\S\ref{sec:DM}) used. 
A difference in metallicity between
the two components is also ruled out by recent spectroscopic data 
(\citeauthor{yg08} \citeyear{yg08}). Here 
we apply the same CMD test as in the previous section to constrain the 
possibility of a difference in $Y$ between the two components as being 
responsible for the well-known bimodal nature of the cluster's HB. 

The result is shown in Figure~\ref{fig:N1851}. While the quality of the 
data is not as high as in the case of M3, one is still able to derive 
some general conclusions. First, the same theoretical ZAHB does appear 
to provide a reasonable description of the lower boundary of the data, 
both for the red and blue HB components~-- which suggests that at least 
some of the stars on the blue HB have the same $Y$ as 
do the red HB stars. Second, there is a predominance of overluminous 
stars on the blue HB, at colors around $(b\!-\!y)_0 \approx 0.05 - 0.15$. 
While these might in principle be interpreted in terms of a moderate 
level of helium enrichment, perhaps of the order of $3\%-4\%$ on average 
(see Fig.~\ref{fig:N1851-Y}), the more straightforward explanation is 
that these stars actually represent the well-evolved progeny of the blue
ZAHB stars that are found at higher temperatures. If so, this would again
suggest that the late stages of HB evolution are somehow significantly 
underestimated by present-day HB tracks, similar to what was previously 
found elsewhere \citep[][and references therein]{mcea01,vc08}. 

Clearly, more work is needed before we are able to conclusively establish
the nature of NGC~1851's bimodal HB and SGB \citep[see also][]{scea08,msea08}.

\section{The Oosterhoff Dichotomy and the Formation of the Milky Way}
Irrespective of our ability to properly model HB stars, we can use 
RR Lyrae stars to derive entirely empirical constraints on the process 
of formation of the Milky Way. In the \citet{sz78} scenario, 
much like in modern $\Lambda$CDM cosmology, one expects galaxies
such as our own to have formed 
by the accretion of ``protogalactic fragments'' that may have resembled
the early counterparts of the Milky Way's present-day dwarf satellite
galaxies. Useful constraints on recent accretion events 
may be posed by the presence of younger stellar populations in several 
of these galaxies \citep{muea96}. Still, in order to probe what 
happened {\em very early on}, we must look at the ancient 
components~-- and RR Lyrae stars are especially useful in that regard 
\citep[e.g.,][]{mc08}. 

\begin{figure}[t]
\begin{center}
  \centerline{
 \includegraphics[width=5.325in]{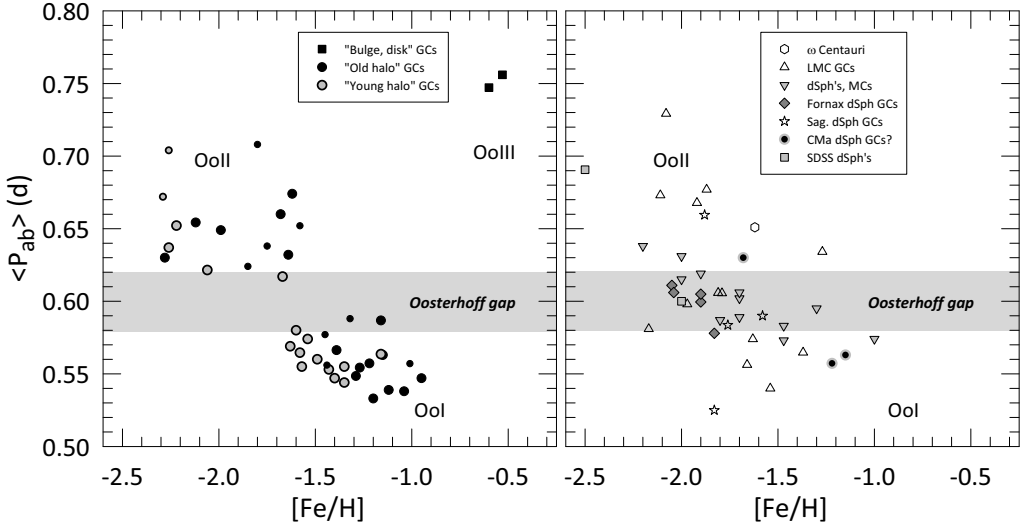} 
 }
 \caption{Distribution of Galactic GC's ({\em left}) and stellar populations 
 associated with neighboring dwarf galaxies ({\em right}) in the average ab-type RR Lyrae
 period $\langle P_{\rm ab}\rangle$ vs. [Fe/H] plane. Galactic GC's are classified into 
 ``bulge/disk,'' ``young halo,'' and ``old halo'' subsystems following \citet{mvdb05}.
 See \citet{mc08} for further details.}
   \label{fig:OOST}
\end{center}
\end{figure}

Are the ancient populations in the Milky Way's dwarf satellites, as traced
by their RR Lyrae pulsators, consistent with the Galactic spheroid having
been built therefrom?  
The answer is provided in Figure~\ref{fig:OOST}, where we compare the
average properties of the fundamental-mode (ab-type) RR Lyrae stars in 
Galactic ({\em left panel}) vs. nearby extragalactic ({\em right panel})
GC's and field populations. While the 
Galactic distribution clearly presents the so-called 
{\em Oosterhoff dichotomy}, with a tendency for systems to clump around
the ``Oosterhoff I'' (OoI) and ``Oosterhoff II'' (OoII) regions
\citep[see also][for the case of halo field stars]{amea08}, 
the opposite happens in the case of nearby extragalactic 
systems, which tend to be preferentially 
{\em Oosterhoff-intermediate}.
This strongly suggests 
that the oldest components of the Galaxy cannot have been formed 
by accretion of even the early counterparts of its present-day
dwarf galaxy satellites. 

As indicated in Figure~\ref{fig:OOST}, at least one of the 
newly discovered SDSS dwarf galaxies \citep[e.g.,][]{vbea06,vbea07}, 
CVn~I, is Oosterhoff intermediate \citep{ckea08}, whereas the 
Bootes dwarf is OoII \citep{mdoea06,ms06}. Unfortunately, some of the 
low-mass SDSS 
galaxies seem to harbor a mere one or two RR Lyrae stars, which makes
it more difficult to assign a conclusive Oosterhoff status to them. 
Indeed, due to statistical fluctuations, and 
since the HB lifetime is of order 100~Myr, it is not entirely clear 
whether the same Oosterhoff types would 
necessarily be inferred for these galaxies if they 
were observed, say, a few hundred Myr in the future (which is very 
little, in terms of Galactic history), when  
these HB stars will have long left the HB phase, to be 
replaced by an entirely new generation of HB stars. 
Still, the present-day properties for several of the RR Lyrae 
stars that are found in these very low-mass galaxies 
do appear to be consistent with an OoII status 
\citep[e.g.,][]{cgea08}.

\section{Conclusions} 
\begin{itemize}
\item HB stars play a central role in the age debate. Still, before we are 
able to predict how (ZA)HB temperature changes with age, 
we must 
properly describe RGB mass loss. 
%settle the issue of how RGB stars lose mass. 

\item In studies of HB morphology, 
it is not sufficient anymore to analyze solely the ``horizontal'' 
HB morphology, meaning the temperature and/or color distribution 
of HB stars: {\em ``vertical''} HB morphology, or the distribution 
of HB stars in luminosity {\em at fixed temperature (or 
color)}, provides us with unique information to help us constrain 
theoretical scenarios for the origin and evolution of these stars. 
Indeed, the available data appears to strongly 
constrain, if not conclusively rule out, the possibility of significant
He enhancements among M3's blue HB stars, while at the same time
suggesting that canonical HB models underestimate the duration of 
the late stages of HB evolution. 

\item Irrespective of our ability
to model them, HB stars~-- and RR Lyrae in 
particular~-- represent invaluable tools to probe into the 
Milky Way's early formation history. 
\end{itemize}

\vskip 0.255cm

\noindent {\bf Acknowledgments.} I would like to warmly thank Gisella Clementini, 
    Frank Grundahl, Bob Rood, Horace Smith, Allen Sweigart, and Aldo Valcarce for 
	useful discussions, comments and suggestions. 
    Support for this work is provided
    by Proyecto Basal PFB-06/2007, by FONDAP Centro de Astrof\'{i}sica 15010003, 
	by Proyecto FONDECYT \#1071002, 
	and by a John Simon Guggenheim Memorial Foundation Fellowship.

\end{document}